# Scalable High-Temperature Superconducting Diodes in Intrinsic Josephson Junctions

Zihan Wei[1,2,+], Youkai Qiao[3,+], Yang-Yang Lyu[1,+,*], Da Wang[3,*], Tianyu Li[1], Leonardo Rodrigues Cadorim[4], Ping Zhang[1,2], Wen-Cheng Yue[1], Dingding Li[1,2], Ziyu Song[1], Zixi Wang[1], Yunfan Wang[1], Milorad V. Milošević[4], Yong-Lei Wang[1,2,5,*], Huabing Wang[1,2,*], Peiheng Wu[1,2]

[1]*Research Institute of Superconductor Electronics (RISE) & Key Laboratory of Optoelectronic Devices and Systems with Extreme Performances of MOE, School of Electronic Science and Engineering, Nanjing University; Nanjing 210023, China*

[2]*Purple Mountain Laboratories; Nanjing 211111, China*

[3]*National Laboratory of Solid State Microstructures & School of Physics, Nanjing University; Nanjing 210023, China*

[4]*COMMIT, Department of Physics, University of Antwerp; Antwerp 2000, Belgium*

[5]*State Key Laboratory of Spintronics Devices and Technologies, Nanjing University; Nanjing 210093, China*

+These authors contributed equally to this work.

*Email: yylyu@nju.edu.cn; dawang@nju.edu.cn; yongleiwang@nju.edu.cn; hbwang@nju.edu.cn

# Abstract

Superconducting diodes, characterized by nonreciprocal supercurrent transport, offer transformative opportunities for ultra-low-power circuits. However, achieving reliable operation at temperatures above liquid nitrogen remains a major challenge, limiting their practical applicability. Here, we present a scalable strategy for high-temperature superconducting diodes based on intrinsic Josephson junctions naturally present in a cuprate superconductor. We demonstrate that strong nonreciprocity arises not only from broken spatial and time-reversal symmetries, but also from enhanced anharmonicity in the current-phase relation, enabled by the atomically thin barrier of the intrinsic junction. The diode efficiency strongly depends on the number of stacked intrinsic junctions, with the highest efficiency occurring in single-junction devices. Notably, these high-temperature superconducting diodes are readily scalable to large arrays, marking a critical step toward practical implementation in energy-efficient computing architectures.

# Main Text

Nonreciprocal charge transport—the ability to preferentially conduct electrical current in one direction—is central to modern electronics, enabling essential components such as rectifiers, voltage regulators, and photodetectors. While well-established in semiconducting p-n junctions, extending nonreciprocity into the superconducting regime has recently sparked intense interest, driven by the potential for next-generation energy-efficient electronic technologies[1-4]. The superconducting diode effect, wherein supercurrent exhibits directional asymmetry, has now been experimentally demonstrated in a variety of superconducting systems, including superlattices[5-7], two-dimensional materials[8-10], and microbridges[11,12]. Among these platforms, the Josephson junction, formed by sandwiching a thin non-superconducting barrier between two superconductors, has emerged as a particularly versatile unit for realizing the superconducting diode effect. Implementations have ranged from planar junctions[13-15], magnetic-atom junctions[16], van der Waals heterostructures[17-19], to superconducting quantum interference devices[20-22] and multiterminal devices[23,24].

The majority of these Josephson diodes rely on low-$T_c$ superconductors, thereby constraining their operating temperature. Achieving superconducting diodes in high-$T_c$ superconductors would significantly expand the operational temperature window, reduce cooling demands, and enable more practical deployment. The high-$T_c$ cuprate superconductor $Bi_2Sr_2CaCu_2O_{8+\delta}$ (BSCCO), with its layered structure, offers a unique platform for such efforts[25-28]. Recent advances have explored high-temperature superconducting diodes using twisted BSCCO flakes to form artificial Josephson junctions[26-28]. However, these devices typically depend on highly specialized exfoliation and stacking techniques, posing challenges for consistent large-scale integration.

The highly anisotropic BSCCO crystal structure (Fig. 1a), with its layered architecture, intrinsically hosts stacks of Josephson junctions—so-called intrinsic Josephson junctions[29]. These naturally formed intrinsic Josephson junctions have been extensively utilized in solid-state terahertz sources[30,31] and detectors[32,33]. Despite this, no superconducting diode effect has been reported in intrinsic Josephson junctions. In this work, we demonstrate a lithography-compatible approach to realize high-temperature Josephson diodes directly from BSCCO intrinsic Josephson junctions. The resulting devices exhibit pronounced nonreciprocal critical currents across a broad temperature range, reaching up to 86 K, thus surpassing the liquid nitrogen threshold and enabling more accessible cryogenic operation. We further demonstrate that the nonreciprocity is tunable via the stack number of intrinsic Josephson junctions. A large diode efficiency exceeding 40% is realized in a single-junction device. These devices exhibit a memory effect, offering nonvolatile control. Most importantly, our approach is readily scalable to large arrays containing hundreds of intrinsic Josephson diodes, paving the way for integrating high-temperature superconducting diodes into scalable electronic and quantum circuits[34].

## High-temperature intrinsic Josephson diodes

The intrinsic Josephson junction devices are created by leveraging the layered lattice structure of BSCCO crystals (Fig. 1a). Each device consists of a BSCCO stack encapsulated between top and bottom superconducting electrodes, both of which are coated with gold films (Fig. 1b). The stacked BSCCO intrinsic Josephson junctions are patterned into a wedge geometry with an inclination angle of 23 degrees. Detailed fabrication procedures are provided in Methods. The current-voltage characteristics measured at 80 K (Figs. 1c and 1d) show well-separated superconducting critical currents, $I_c^+$ and $I_c^-$, measured under positive and negative currents, respectively, confirming the realization of a high-temperature superconducting diode. The diode polarity can be reversed by altering the direction of the applied out-of-plane magnetic field, $H_z$ (Figs. 1c and 1d). When driven by a square-wave current excitation, the device steadily switches between zero-resistance and normal-resistance states (Fig. 1e). Our temperature-dependent measurements reveal that the device operates as a superconducting diode up to 86 K (Fig. 1f), representing the highest operation temperature reported to date for superconducting diodes.

## Microscopic mechanism of intrinsic Josephson Diodes

In general, the superconducting diode effect is induced by simultaneously breaking spatial-inversion and time-reversal symmetries[35]. In our intrinsic Josephson diode, spatial-inversion symmetry is broken by the wedge-shaped geometry of the device (Fig. 1b). Time-reversal symmetry is typically broken by applying a magnetic field *perpendicular* to the current direction[36,37]. In intrinsic Josephson junctions, the Josephson supercurrent flows along the out-of-plane (z-axis) direction of the BSCCO crystal. Therefore, one would conventionally expect the superconducting diode effect to be induced by an *in-plane* magnetic field, which is perpendicular to the Josephson supercurrent. Contrary to this expectation, our field-orientation experiments unambiguously show that only the out-of-plane component of the magnetic fields—parallel to the Josephson supercurrent and not traditionally associated with time-reversal symmetry breaking—dominates the superconducting diode effect, while in-plane fields have a negligible influence (Extended Data Fig. 1). Recent studies have suggested that time-reversal symmetry can be broken in twist-angle van der Waals Josephson junctions formed between BSCCO crystal flakes[26-28]. However, our intrinsic junctions do not involve any twist angle, implying that a distinct mechanism is responsible for the observed Josephson diode effects in our devices.

To uncover the microscopic mechanism of the intrinsic Josephson diodes, we model our device using the Lawrence-Doniach formalism[38], treating the interlayer Josephson current (out-of-plane) and intralayer supercurrent (in-plane) simultaneously as shown in Fig. 2a. The supercurrent flows in from one corner of the top layer and flows out from the opposite corner of the bottom layer (Fig. 2a). We have derived a modified self-consistent current-phase relation along each current channel (Extended Data Figs. 2a-c):

$$i = \sin(\phi - \frac{2e}{\hbar}\int \vec{A}\cdot d\vec{l} - \kappa i) \quad (1)$$

Here, $i$ is a dimensionless current, $\frac{2e}{\hbar}\int \vec{A}\cdot d\vec{l}$ represents the phase shift induced by the out-of-plane magnetic field, associated with time-reversal symmetry breaking[39]. $\kappa$ is a dimensionless coefficient characterizing the anharmonicity for each current channel[40], which describes the deviation of the current-phase relation from a simple sinusoidal form (details can be found in Methods). Figure 2b shows the calculated superconducting diode effect, which is consistent with the experimental observations in Fig. 1d.

We further analyzed the spatial distribution of both the in-plane intralayer supercurrent and the out-of-plane Josephson supercurrent (Fig. 2c). The results reveal that the out-of-plane magnetic field induces a nonreciprocal modulation of the in-plane supercurrent in the top and bottom $CuO_2$ layers, which in turn leads to a nonreciprocal distribution of the out-of-plane Josephson supercurrent, thereby breaking time-reversal symmetry. Notably, finite values of the anharmonicity coefficient $\kappa$ are essential for the superconducting diode effect (Extended Data Figs. 2d-f). Our simulations indicate that when $\kappa$ is zero, no diode effect occurs, even in the presence of time-reversal symmetry breaking. In conventional Josephson tunnel junctions, the anharmonicity is typically very weak[41], which results in negligible superconducting diode effects. In our modeling, the anharmonicity coefficient $\kappa$ scales approximately as $L^2/d\xi$, where $L$ is the length of each current channel from input to output leads, $d$ is the barrier thickness of the intrinsic Josephson junctions, and $\xi$ is the planar superconducting coherence length (see Methods). Therefore, the anharmonicity $\kappa$ is intrinsically determined by the device geometry and electrode configuration. The atomic-scale barrier thickness in BSCCO ($d\rightarrow 0$) leads to a significantly enhanced $\kappa$, thereby producing a prominent superconducting diode effect in intrinsic Josephson junctions.

**Junction number-dependent intrinsic Josephson diodes**

The current-voltage characteristics of our device indicate that there are three intrinsic junctions stacked in series[42] (Extended Data Fig. 3). The theoretical model described above can be extended to an $N$-intrinsic Josephson junction stack (inset of Fig. 2d). The current-phase relationship for each current channel in the $N$-junction stacks becomes:

$$i = \sin\left[\frac{1}{N}\left(\phi - \frac{2e}{\hbar}\int \vec{A}\cdot d\vec{l} - \kappa i\right)\right]. \qquad (2)$$

This equation shows that the anharmonicity weakens as the junctions increase. In Fig. 2d, we show the calculated superconducting diode efficiency $\eta$, defined as $\eta = (I_c^+ - |I_c^-|) / (I_c^+ + |I_c^-|) \times 100\%$, as a function of the number of stacked junctions. It clearly predicts a monotonic decay of $\eta$ with increasing junctions.

To experimentally validate this behavior, we fabricated two more devices, which contain approximately 14 and 17 intrinsic Josephson junctions, respectively (the corresponding current-voltage characteristics are shown in Extended Data Fig. 4). Figures 2e-2g show the field-dependent diode efficiency for all three devices with different numbers of junctions. The results clearly indicate that the diode efficiency is gradually enhanced in devices with fewer junctions, in agreement with our theoretical predictions. These findings further highlight the critical role of anharmonicity in enabling the superconducting diode effect in intrinsic Josephson junctions.

**Enhanced performance in surface intrinsic Josephson diodes**

As reducing the number of junctions enhances superconducting diode efficiency (Fig. 2d), one would expect the maximum diode efficiency $\eta$ in a single intrinsic Josephson junction. However, precisely fabricating a single Josephson junction remains a considerable technical challenge. Fortunately, the surface intrinsic Josephson junction, formed between the two uppermost $CuO_2$ bilayers of the BSCCO crystal (top panel of Fig. 3a), provides a reliable route to access a single intrinsic Josephson junction[43-45]. We fabricated a wedge-shaped surface intrinsic Josephson junction device (bottom panel of Fig. 3a). The detailed fabrication procedures are provided in Methods and Ref. 45.

The surface junction device exhibits a pronounced superconducting diode effect (Fig. 3b). Unlike the multi-junction diodes, which exhibit a relatively wide distribution of $I_c$ (Extended Data Fig. 3d), the surface intrinsic Josephson diode shows excellent reliability with sharp and reproducible switching currents under repeated operation (Fig. 3b). Notably, the diode efficiency reaches values up to 40% (Fig. 3c), substantially higher than those in multi-junction diodes (Figs. 2e-2g).

Our surface Josephson diodes display a unique memory effect that enables nonvolatile switching and reversal of the diode's nonreciprocity. When sweeping the magnetic field in a low-field region (Fig.

3c), the device exhibits an antisymmetric field response between positive and negative current configurations, with no superconducting diode effect at the zero magnetic field—that is $I_c^+(H_z=0)=I_c^-(H_z=0)$. Interestingly, upon applying a large magnetic field ($|H_z|>60$ Oe) to magnetize the device, a memory effect emerges (Fig. 3d): the values of $I_c^+$ and $I_c^-$ near zero magnetic field become sensitive to the device's magnetization history. Consequently, it displays a superconducting diode effect at the zero magnetic field, as shown in Fig. 3e, with its diode polarity nonvolatilely determined by the direction of the initializing magnetic field. Importantly, the zero-field superconducting diode efficiency remains notable, with $\eta(H_z=0)\approx 20\%$. Figure 3f demonstrates rectification of an AC signal to a DC output at zero field, with nonvolatile and reversible functionality.

The zero-field superconducting diode (or memory effect) most likely originates from the trapped Abrikosov vortices in the BSCCO crystals beneath the surface junction, which breaks time-reversal symmetry even in the absence of an applied field. The magnetic memory of the superconducting diode can be erased by warming the device above its superconducting transition temperature (Extended Data Fig. 5), consistent with the behavior of trapped magnetic flux. These results highlight both the robustness and exceptional performance of surface intrinsic Josephson diodes.

## Scalable high-temperature superconducting diodes

Taking advantage of the top-down fabrication method and the precise control over the size and geometry of our lithographically patterned intrinsic Josephson diodes, we demonstrate the feasibility of creating a large array of high-temperature superconducting diodes. As shown in Fig. 4a, we fabricated a large-scale array consisting of 201 serially connected surface intrinsic Josephson diodes. Transport measurements of the selected device reveal highly consistent, reliable, and nonvolatilely programmable superconducting diode behavior at zero magnetic field (Figs. 4b and Extended Data Fig. 6), validating the scalability and reproducibility of high-temperature intrinsic Josephson diodes for technological applications.

## Conclusions

Our high-temperature superconducting diodes based on intrinsic Josephson junctions exhibit all the essential properties for practical and large-scale applications, including high-temperature operation, significant diode efficiency, nonvolatile reconfigurability (memory effect), and—most importantly—scalability to large arrays. By leveraging lithographically defined device geometry and the controllable number of stacked junctions, our approach overcomes material degradation challenges and achieves robust and reproducible superconducting diodes. Both theoretical modeling and experiments reveal that geometry-induced anharmonicity plays a key role in enabling strong nonreciprocity, with the superconducting diode effect significantly enhanced as the number of intrinsic Josephson junctions is reduced—culminating in optimal performance in single junction devices. This design strategy provides a versatile platform for future optimization, such as enhancing anharmonicity through tailored geometries and incorporating artificial pinning centers, including magnetic nanostructures[46] or nanoholes[47], into the intrinsic Josephson junction architecture for a more deterministic and higher-performance superconducting diode. Our results clarify the fundamental understanding of superconducting diode phenomena in high-$T_c$ superconductors. They also advance the development of scalable, on-chip superconducting circuits[48] and unique quantum device architectures[49] based on intrinsic Josephson phenomena in high-$T_c$ superconductors.

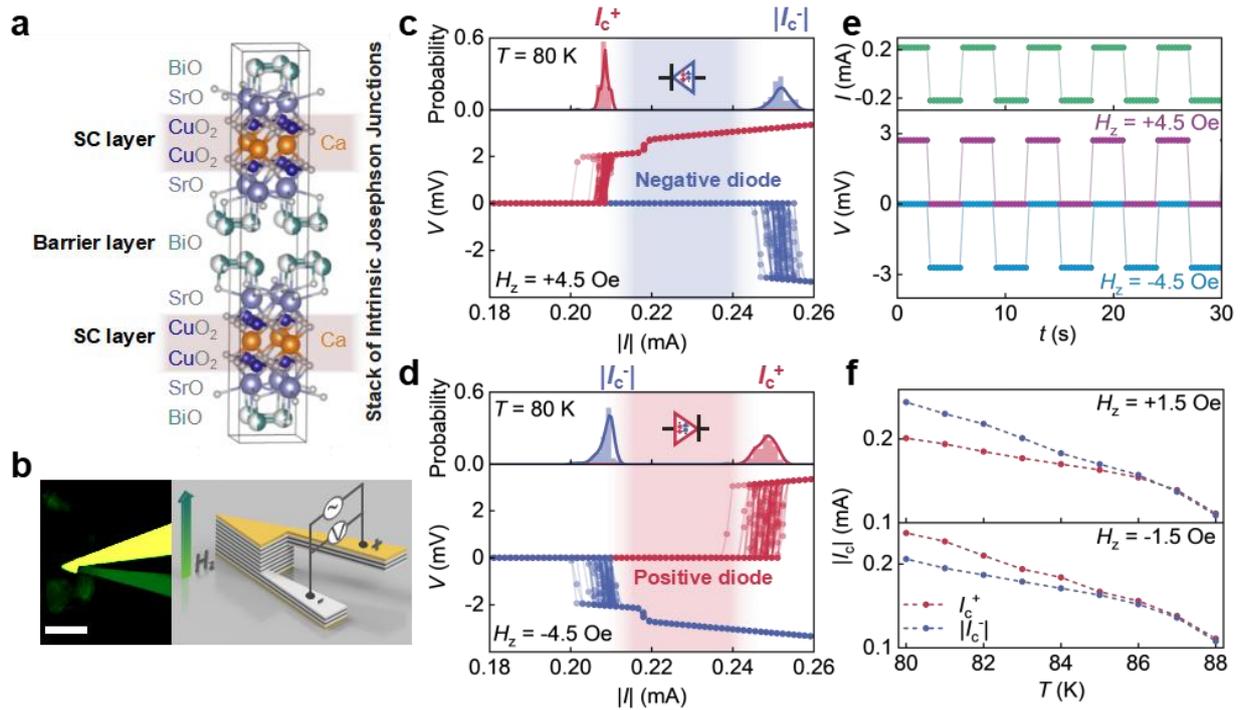

**Fig. 1 | High-temperature intrinsic Josephson diodes. a,** Crystal structure of BSCCO, composed of alternating superconducting $CuO_2$ bilayers (SC layer) and insulating SrO-BiO barriers, which naturally form intrinsic Josephson junctions. **b,** Optical image and schematic diagram of the intrinsic Josephson diode. Scale bar: 30 μm. **c, d,** Nonreciprocal supercurrent transport at 80 K under the magnetic field of +4.5 Oe (**c**) and -4.5 Oe (**d**), respectively. The lower panels display 100 repeated current-voltage sweeps with positive (red) and negative (blue) currents. The upper panels show statistical distributions of the superconducting critical currents $I_c^+$ and $|I_c^-|$. **e,** Half-wave rectification at 80 K for $H_z$=+4.5 Oe (purple) and $H_z$=-4.5 Oe (blue). When excited by a 0.22 mA square-wave current (green), the device exhibits a clear polarity reversal upon inversion of the magnetic field direction. **f,** Temperature dependence of nonreciprocity. The Josephson diode effect, quantified by the asymmetry in critical currents $I_c^+ \neq |I_c^-|$, remains robust up to 86 K.

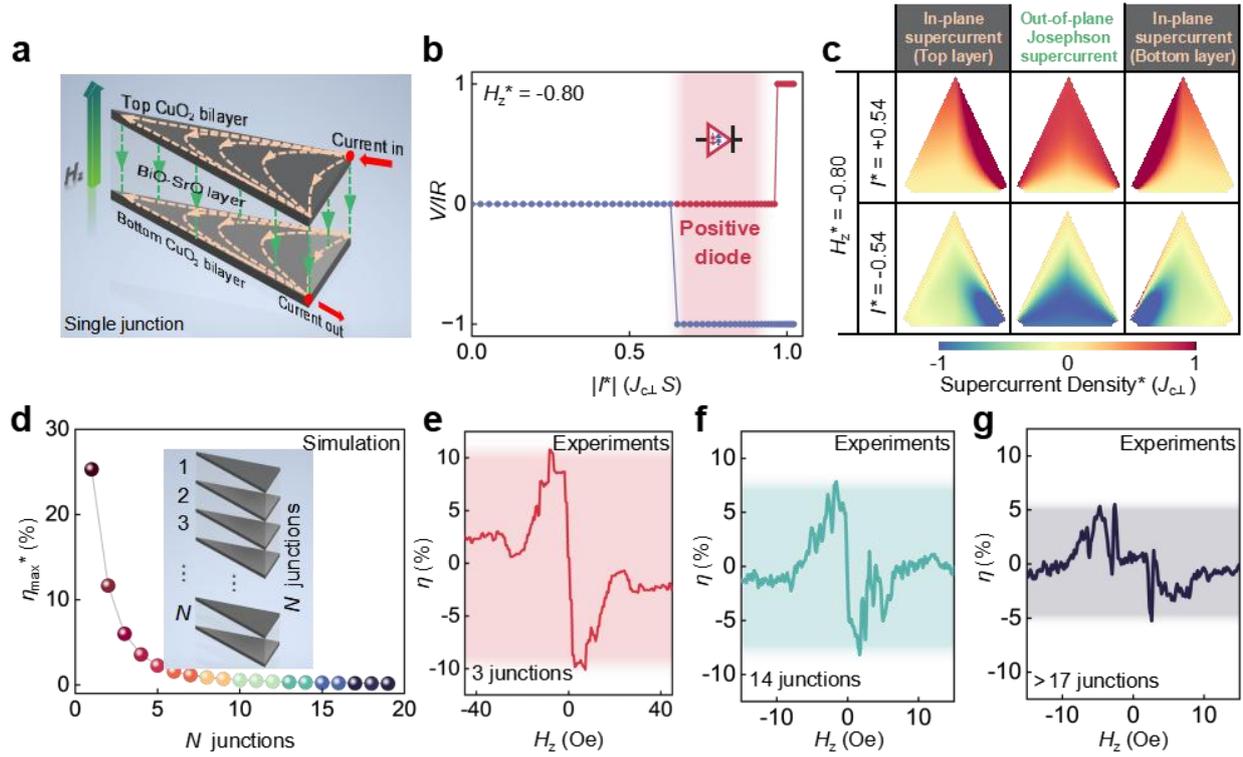

**Fig. 2 | Mechanism and junction-number dependence of intrinsic Josephson diodes. a,** Schematic of the theoretical model incorporating both in-plane supercurrents in $CuO_2$ bilayers (orange arrows) and out-of-plane interlayer Josephson supercurrents (green arrows). **b,** Calculated current-voltage characteristics. The critical currents exhibit strong asymmetry under the magnetic field (positive bias: red; negative bias: blue). The magnetic field is normalized by $\Phi_0/\mu_0 S$, current by $J_{c\perp}S$, and the voltage spans from 0 (superconducting state) to ±1 (resistive state). Here, $\Phi_0$ is the magnetic flux quantum, $\mu_0$ is the vacuum permeability, $S$ is the junction area, and $J_{c\perp}$ is the Josephson critical supercurrent density. **c,** Spatial distributions of supercurrent density under an out-of-plane field, highlighting broken time-reversal symmetry. **d,** Calculated diode efficiency as a function of junction number. **e-g,** Experimental diode efficiencies under magnetic fields, extracted from $I_c^+/|I_c^-|$ curves in **Extended Data Figs. 1d, 4b, and 4d,** respectively. The number of junctions, labeled in each graph, is identified from the corresponding current-voltage characteristics in **Extended Data Figs. 3b, 4a, and 4c**.

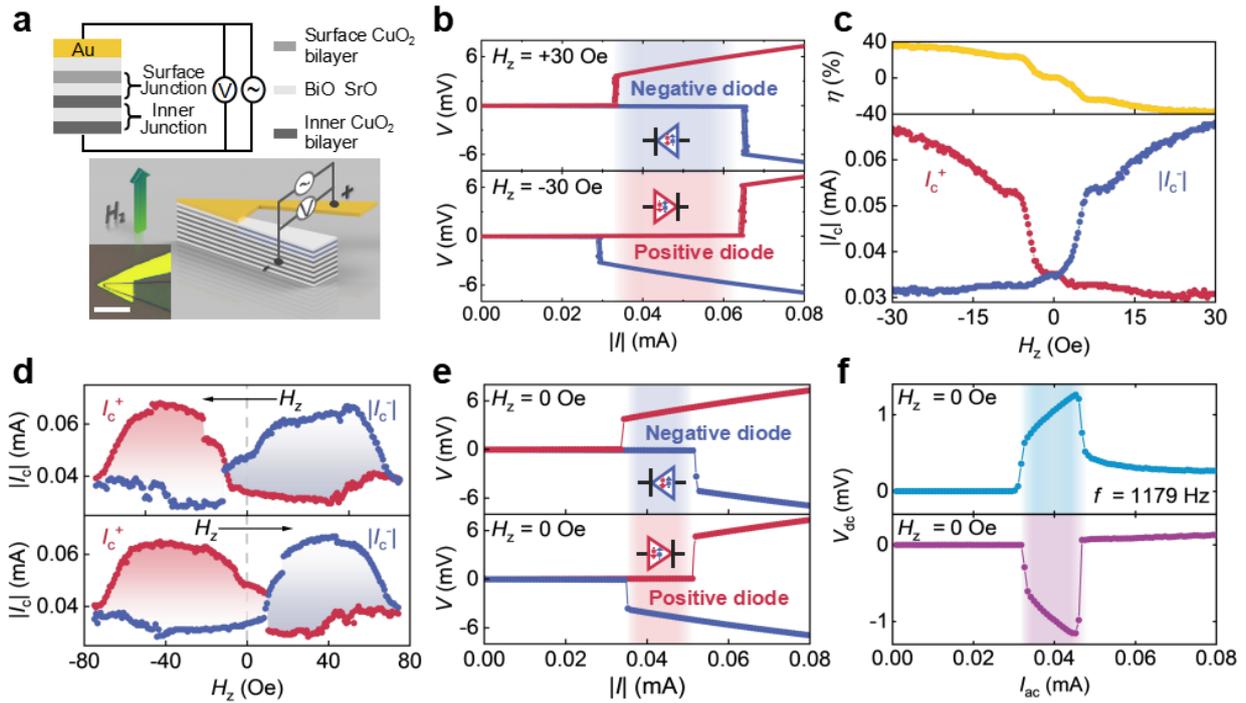

**Fig. 3 | Surface intrinsic Josephson diode comprising a single-junction. a,** Schematic of the device structure. Superconductivity in the topmost surface CuO$_2$ layer, directly contacted by the gold electrode, is slightly suppressed, resulting in a reduced critical current compared to inner junctions. The lower illustration shows the device layout. BSCCO regions unprotected by gold film (blue and white stacks) have degraded. The inset displays an optical micrograph of the device. Scale bar: 30 μm. **b,** Forty repeated current-voltage sweeps, demonstrating a robust and reproducible Josephson diode effect at 3 K. **c,** Field-dependent critical currents and corresponding diode efficiency obtained in the low-field range. **d,** Memory effect under extended magnetic fields. Reversing the field sweep direction switches the diode polarity at zero field (gray dashed line). **e,** Current-voltage characteristics of superconducting diode effects at zero field. The diode polarity is programmed by initialization fields of $H_z$=+75 Oe (upper panel) and $H_z$=-75 Oe (lower panel), respectively. **f,** Superconducting rectification measured under an AC excitation of 1179 Hz and at zero field.

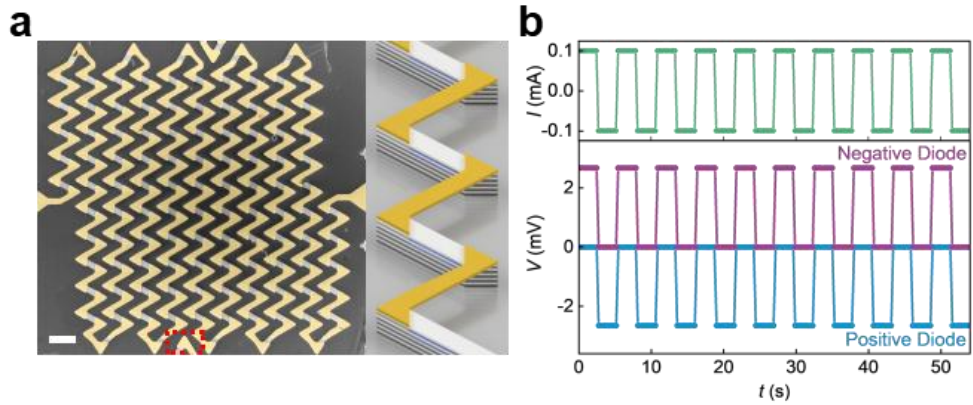

**Fig. 4 | Scalable integration of Josephson diode with surface junctions. a,** Pseudocolored SEM image (left panel) of an array containing 201 serially connected surface intrinsic Josephson diodes. Scale bar: 50 μm. The right panel illustrates the schematic of the diode array configuration. Gold electrodes are shown in yellow, while degraded BSCCO regions are depicted as blue and white stacks. **b,** Zero-field half-wave rectification. A square-wave current excitation (green), applied to the diode highlighted in the red box in (**a**), produces rectified voltage outputs at zero field for both negative (purple) and positive (blue) diode polarities.

## Methods

**Fabrication of intrinsic Josephson junction device**

BSCCO single crystals were grown using the floating-zone method. Intrinsic Josephson junction devices were fabricated via a double-sided patterning technique[32] (Extended Data Fig. 7). Crystals were mounted on sapphire substrates using epoxy, with the *c*-axis oriented perpendicular to the substrate. A 50-nm-thick gold film was immediately deposited onto the freshly cleaved surface (a). A wedge-shaped mesa was subsequently defined using standard photolithography and ion milling, producing a structure with a typical thickness of ~300 nm (b). Additional ion milling was applied to etch step structures at the wedge, with an etching depth slightly less than 300 nm, to define one side of the device electrode (c). The entire surface was then encapsulated in epoxy, and a second sapphire substrate was applied on top (d). After curing at elevated temperature, the sample was mechanically split (e) and flipped onto the new substrate (f). A second 50-nm-thick gold layer was deposited on the newly exposed surface, and the opposite electrode was patterned through a third round of photolithography (g). The final number of junctions in each device was controlled by adjusting the final etching depth of ion milling (h).

**Fabrication of surface intrinsic Josephson junction device**

The top BSCCO electrode in conventional intrinsic Josephson junction devices was replaced by a gold electrode, deposited via *in-situ* high-vacuum evaporation at room temperature (Extended Data Fig. 8). A small piece of BSCCO crystal was mounted onto a sapphire substrate using epoxy, with its *c*-axis perpendicular to the surface. The sample was then transferred into a high-vacuum evaporator, where a piece of Scotch tape was attached to the shutter to enable *in-situ* cleavage of the crystal. Once the chamber was preheated to the desired temperature, the shutter was opened to cleave the crystal, immediately followed by the deposition of a 200-nm-thick gold film (a) at a deposition rate of ~18 nm/s. A smooth region on the cleaved surface was selected, and wedge-shaped BSCCO stacks were patterned using conventional photolithography and ion milling (b). A $SiO_x$ insulating layer was subsequently deposited and lifted off in acetone, forming an insulating layer with thickness matched to the surrounding crystal (c). A 100-nm-thick gold layer was deposited and patterned using photolithography (d). Finally, wet etching ($KI/I_2$) was used to define electrodes while minimizing damage to the underlying BSCCO layers (e). The surface intrinsic Josephson junctions are located in the wedge tip region, characterized by a top angle of 23 degrees.

**Transport measurements**

Transport measurements were performed in cryogen-free cryostats equipped with three-axis superconducting magnets (Cryomagnetics, Inc.) for intrinsic Josephson junction devices and a GM refrigerator (Sumitomo Corp.) with a custom-built coil for surface intrinsic Josephson junction devices. DC measurements utilized a homemade current source and a low-noise voltage amplifier, powered by a 12 V lead-acid battery to suppress noise. Data acquisition was controlled via an NI PCI-6221 data acquisition card. For AC measurements, a Keithley 6221 current source and 2182A nanovoltmeter were employed. Intrinsic Josephson junction devices were configured for four-terminal measurements, while surface intrinsic Josephson junction devices used a three-terminal setup, with the top gold layer serving both current and voltage leads. The contact resistance between the gold electrode and the uppermost $CuO_2$ layer was typically lower than 1 Ω.

**Lawrence-Doniach simulation**

To model the behavior of layered superconductors, we adopted the Lawrence–Doniach framework, which extends beyond the conventional anisotropic Ginzburg–Landau theory by treating the interlayer coupling as Josephson tunneling. In this model, the in-plane supercurrent density within the $n$-th layer is governed by the London equation:

$$\vec{J}_{//}^{n} = \rho_s \left( \frac{\hbar}{2e} \nabla \phi_n - \vec{A} \right), \qquad (3)$$

while the Josephson supercurrent density between the $n$-th and $(n+1)$-th layers is described by the Josephson relation:

$$J_{\perp}^{n+1,n} = J_{c\perp} \sin(\phi_{n+1} - \phi_n), \qquad (4)$$

where $\rho_s$ is the in-plane superfluid density, $J_{c\perp}$ is the critical Josephson current density, $\phi_n$ is the phase of the $n$-th layer, and $\vec{A}$ is the vector potential.

**Temperature dependence of intrinsic Josephson diode**

Intrinsic Josephson junction devices exhibit a superconducting transition temperature of approximately 90 K (Extended Data Fig. 3a). At a low temperature of 3 K, the distributions of the statistically evaluated critical currents $I_c^+$ and $|I_c^-|$, obtained from 1000 repeated current-voltage curves, are relatively broad (Extended Data Fig. 3c). Under zero magnetic field, the distributions of $I_c^+$ and $I_c^-$ nearly coincide, indicating the absence of a superconducting diode effect. An out-of-plane magnetic field ($H_z$) leads to a clear separation of $I_c^+$ and $I_c^-$ into two peaks, revealing the emergence of the superconducting diode effect. The partial overlap of the $I_c^+$ and $|I_c^-|$ distributions limits switching accuracy, consistent with behaviors previously reported in Josephson diode based on twist-angle BSCCO flakes[27]. Temperature-dependent measurements (Extended Data Fig. 3d) show a gradual narrowing of the overlap region as temperature increases, with full separation above 65 K, thereby enabling reliable high-temperature operation of the Josephson diodes.

**Magnetic field orientation dependence of intrinsic Josephson diode**

To investigate the dependence of the intrinsic Josephson diode effect on magnetic field orientation, we characterized our device under three-dimensional magnetic field configurations (Extended Data Fig. 1). Extended Data Fig. 1b shows the critical currents measured under magnetic fields $H_r$ applied at three different directions ($\theta = 30°$ and $\varphi = 0°$, 45° and 90°). Asymmetric values of $I_c^+$ and $I_c^-$ were observed in each case within [-20 Oe, +20 Oe]. The near-complete overlap of the three $I_c$-$H_r$ curves indicates a negligible influence from the in-plane field components ($H_x/H_y$). To clarify the role of the out-of-plane field component, we measured the critical currents in the x–z field plane ($\varphi = 0°$) while varying $\theta$ (Extended Data Fig. 1c). Despite changes in $\theta$ from 0° to 45°, the overall shapes of the $I_c$-$H_r$ curves remained essentially unchanged. When plotted as a function of the out-of-plane field component (Extended Data Fig. 1d), all curves collapsed onto a single trace, confirming that the out-of-plane field is the dominant component driving the intrinsic Josephson diode effect. This behavior is consistent with the two-dimensional nature of the layered BSCCO crystals.

**Transport properties of surface intrinsic Josephson diode**

Surface intrinsic Josephson junctions exhibit weaker interlayer coupling compared to inner intrinsic Josephson junctions, resulting in lower critical currents (Extended Data Fig. 9b) and reduced superconducting transition temperatures (Extended Data Fig. 9a). This enables selective probing of surface intrinsic Josephson junction dynamics at low bias. Over a broad current range, the device displays typical current-voltage characteristics of intrinsic Josephson junctions (Extended Data Fig. 9b). At low currents, a clear hysteresis—indicative of a single Josephson junction—is observed (inset of Extended Data Fig. 9b).

When driven by a square-wave current with amplitude between $I_c^+$ and $|I_c^-|$, the device maintained stable diode operation for over 10,000 cycles (Extended Data Fig. 9c), demonstrating both robust performance and excellent rectification behavior. Furthermore, the zero-field superconducting diode effect was retained following exposure to large field magnetization. The onset threshold for such field-programmed diode behavior was determined by systematically varying the applied $H_z$ field (Extended Data Fig. 9d).

**Simulation model of intrinsic Josephson junction**

We first consider a current channel within a single intrinsic Josephson junction (Extended Data Figs. 2a, 2b). The supercurrent enters from the top-right and exits from the bottom-left corner. Due to the non-uniformity of in-plane current density, the cross-sectional width $b_l$ of the channel varies along the propagation distance $l$. Neglecting the vector potential for simplicity, we assume that the phase difference $\delta_l$ between the two superconducting layers remains constant, denoted as $\delta$, a simplifying approximation validated by numerical calculations (Extended Data Fig. 10). The total current $I$ flowing through the channel is the sum of vertical supercurrents:

$$I = \int J_{c\perp} \sin\delta \; b_l dl = J_{c\perp} S \sin\delta, \tag{5}$$

where $S = \int b_l dl$ is the total area of each layer.

On the other hand, the total current can also be calculated by summing the in-plane supercurrents in both layers:

$$\frac{I}{b_l d} = \frac{\hbar}{2e}\rho_s(\nabla\phi_1 + \nabla\phi_2), \tag{6}$$

where $\phi_1, \phi_2$ are phases for the two layers, $d$ is the vertical inter-layer displacement. Taking a line integral along the path $l$, we obtain

$$\frac{I}{d}\int \frac{dl}{b_l} = \frac{\hbar}{2e}\rho_s(\phi + \phi - \delta - \delta) = \frac{\hbar}{e}\rho_s(\phi - \delta), \tag{7}$$

from which we obtain $\delta = \phi - \left(\frac{e}{\hbar d \rho_s}\int_l b_l^{-1}\right)I$. Substituting this into Eq. (5) gives the current-phase relation:

$$i = \sin(\phi - \kappa i), \tag{8}$$

where $i = \frac{I}{J_{c\perp}S}$ is the dimensionless current and $\kappa$ denotes a dimensionless anharmonicity parameter defined as:

$$\kappa = \frac{e}{\hbar}\frac{J_{c\perp}S}{d\rho_s}\int_l \frac{dl}{b_l} = \frac{1}{3\sqrt{3}}\frac{L^2}{d\xi_{/\!/}}\frac{J_{c\perp}}{J_{c/\!/}}\langle b_l\rangle\langle b_l^{-1}\rangle, \tag{9}$$

where we use the Ginzburg–Landau expression for the in-plane critical current density $J_{c/\!/} = \frac{1}{3\sqrt{3}}\frac{\hbar}{e}\frac{\rho_s}{\xi_{/\!/}}$, $L$ is the channel length, $\langle b_l\rangle$ and $\langle b_l^{-1}\rangle$ are average values of $b_l$ and $b_l^{-1}$ along the channel.

In the presence of a vertical magnetic field, the current-phase relationship within a current channel can be generalized by replacing the phase difference $\phi$ with its gauge-invariant form $\left(\phi - \frac{2e}{\hbar}\int \vec{A}\cdot d\vec{l}\right)$:

$$i = \sin\left(\phi - \frac{2e}{\hbar}\int \vec{A}\cdot d\vec{l} - \kappa i\right), \tag{10}$$

It gives anharmonic current-phase relationship for nonzero $\kappa$. The anharmonicity is enhanced by enlarging $\kappa$ (Extended Data Fig. 2c). From the definition of $\kappa$, it is evident that it can be significantly enhanced by increasing the factor $\frac{L^2}{d\xi_{/\!/}}$.

The total current $I_{\text{tot}}$ is the sum of all the channels (denoted by $k$) $I_{\text{tot}}(\phi) = \sum_k i_k(\phi)$. By varying the phase difference $\phi$ between the two leads, the maximal and minimal values of the total current $I_{\text{tot}}(\phi)$ define the critical current $I_c^+$ and $I_c^-$, respectively. In general, $I_{\text{tot}}(\phi) = \sum_n I_{cn}\sin(n\phi + \delta_n)$. As shown in Extended Data Fig. 2d, the introduction of anharmonicity leads to direction-dependent asymmetry in the critical current under a magnetic field. The critical current difference that reverses with the magnetic field polarity is a direct manifestation of the diode effect. If all the anharmonic terms with $n > 1$ are zero, we must always have $|I_c^+| = |I_c^-|$, hence, without the diode effect (Extended Data Fig. 2e), despite the time-reversal symmetry breaking induced by magnetic fields (Extended Data Fig. 2f). Therefore, anharmonicity is a necessary condition for the Josephson diode effect. This mechanism is further visualized in Video S1, which demonstrates the dynamic in-plane and interlayer current density distributions under negative magnetic field. The magnetic field modulation of current flow, influenced by anharmonicity, leads to $I_c^+ > |I_c^-|$. In contrast, Video S2 shows that without a magnetic field, the current remains symmetric.

**Extension to multi-junctions**

The model generalizes naturally to an $N$-junction stack. Assuming that the vertical phase differences are uniform $\delta$ across all junctions, the in-plane current in middle layers are all zero except top and bottom layers, due to the current conservation. Therefore, phases of different layers from top to bottom are: $\phi_l, \phi_l + \delta, \phi_l + 2\delta, \cdots, \phi_l + N\delta$. Following a similar derivation to the single junction case, the current-phase relation within a current channel generalizes to:

$$i = \sin\left[\frac{1}{N}\left(\phi - \frac{2e}{\hbar}\int \vec{A}\cdot d\vec{l} - \kappa i\right)\right], \tag{11}$$

Thus, increasing the number of junctions $N$ reduces the anharmonicity, which explains the experimentally observed suppression of Josephson diode effect in devices with more junctions.

**Numerical simulations of triangular-shaped device**

We numerically simulated the Lawrence-Doniach model for a triangular-shaped $N$-junction stack. Ideally, both current and phase distributions should be determined by minimizing the total free energy. However, for the complex geometry of triangular multi-junctions, full finite-element simulations are computationally demanding. At this stage, we therefore adopted a simplified model, assuming that the current follows along multiple bilayer channels (Fig. 2a). The device can be regarded as a parallel connection of these current channels, all sharing a common input phase (set to zero) and output phase $\phi$.

For the numerical calculation, each in-plane path is discretized into $M$ slices with length $\Delta = L/M$. The phases of the two layers are denoted as $\phi_{n,l}$ ($n = 1, 2, \cdots N+1$; $l = 0, 1, 2, \cdots M$), respectively. Then, the in-plane current can be expressed as:

$$\frac{I_{//,n,l\to l+1}}{db_l} = \frac{\hbar}{2e}\rho_s \frac{\phi_{n,l+1} - \phi_{n,l} - \frac{2e}{\hbar}\vec{A}_l\cdot\vec{\Delta}_l}{\Delta}, \tag{12}$$

and the vertical current is:

$$I_{\perp, n \to n+1, l} = J_c b_l \Delta \sin\left(\phi_{n+1, l} - \phi_{n, l}\right), \tag{13}$$

In practice, we fix $\phi_{1,0} = 0$, and choose $\phi_{n, 0}(n > 1)$ as variational parameters to minimize an objective function $O\left(\{\phi_{n, 0}\}\right) = |I_{//, 1, M \to M+1}| + |I_{//, 2, M \to M+1}| + \cdots + |I_{//, N, M \to M+1}|$. After optimization, $\phi = \phi_{N+1, M}$ is extracted to obtain the current-phase relation $I(\phi)$ for this channel.

We have performed numerical calculations for $N = 1, 2, 3$, respectively. We found the results of current-phase relation for each channel $I_{N, k}(\phi)$ can be described by Eq. (11) quite well (Extended Data Fig. 10). Based on this observation, we only need to obtain the anharmonicity parameter $\kappa$ for each channel for $N = 1$, and then apply it to any $N > 1$ following Eq. (13) directly. In this way, we can predict the diode coefficient for all multi-junction stacks.

# End notes

## Acknowledgments

We thank Xianjing Zhou for key discussions. We acknowledge financial support by the National Key R&D Program of China (Grant Nos. 2021YFA0718802(H.W.), 2022YFA1403201(D.W.), 2024YFA1408100 (D.W.)), the Innovation Program for Quantum Science and Technology(Grant Nos. 2024ZD0301300(Y.-L.W.)), the National Natural Science Foundation of China (Grant Nos. 62288101(H.W.), 62274086(Y.-L.W.), 12274205(D.W.), 62401647(Z.W.), 62101243(Y.-Y.L.)), Jiangsu Key Laboratory of Advanced Techniques for Manipulating Electromagnetic Waves, Jiangsu Provincial Natural Science Fund (Grant No. BK20240290(Z.W.)), Frontier Technologies R&D Program of Jiangsu (Grant No. BF2024058(Z.W.)), Jiangsu Outstanding Postdoctoral Program (Y.-Y.L. and Z.W.), Research Foundation-Flanders (FWO) (L.R.C., M.V.M.), EU-COST Action CA21144 SUPEROUMAP (L.R.C., M.V.M.) and U.S. Army Research Office W911NF-24-1-0145 (M.V.M.).

## Author contributions

Y.-L.W. and H.W. conceived and designed the study. Z.W. fabricated the devices with the assistance of Z.S., D.L. and Y.W.. Y.-Y.L. performed the transport measurements with the help of Z.W., T.L., P.Z., W.C.Y. and Z.W.. Y.Q. and D.W. contributed theoretical considerations and the model calculations. L.R.C. and M.V.M. provided theoretical support and discussion. Y.-Y.L., Z.W. and D.W. collected the data and wrote the original draft. Y.-L.W. reviewed and edited the manuscript with input from all coauthors. Y.L.W., H.W. and P.W. supervised the study.

## Competing interests

The authors declare no competing interests.

## Additional information

Supplementary Information is available for this paper.

Correspondence and requests for materials should be addressed to Y.-Y.L.

**Supplementary Information**

**Video S1 | Spatial distribution of supercurrent densities under a finite $H_z$ field.** The top six color maps display the spatial distributions of in-plane and Josephson supercurrents under opposite bias currents ($I^*=\pm0.54$) with a fixed out-of-plane magnetic field ($H_z^*=-0.80$). The bottom graph shows the device resistance as a function of bias current.

**Video S2 | Spatial distribution of supercurrent densities at zero field.** The top six color maps display the spatial distributions of in-plane and Josephson supercurrents under opposite bias currents ($I^*=\pm0.54$) at zero field ($H_z^*=0$). The bottom graph shows the device resistance as a function of bias current.

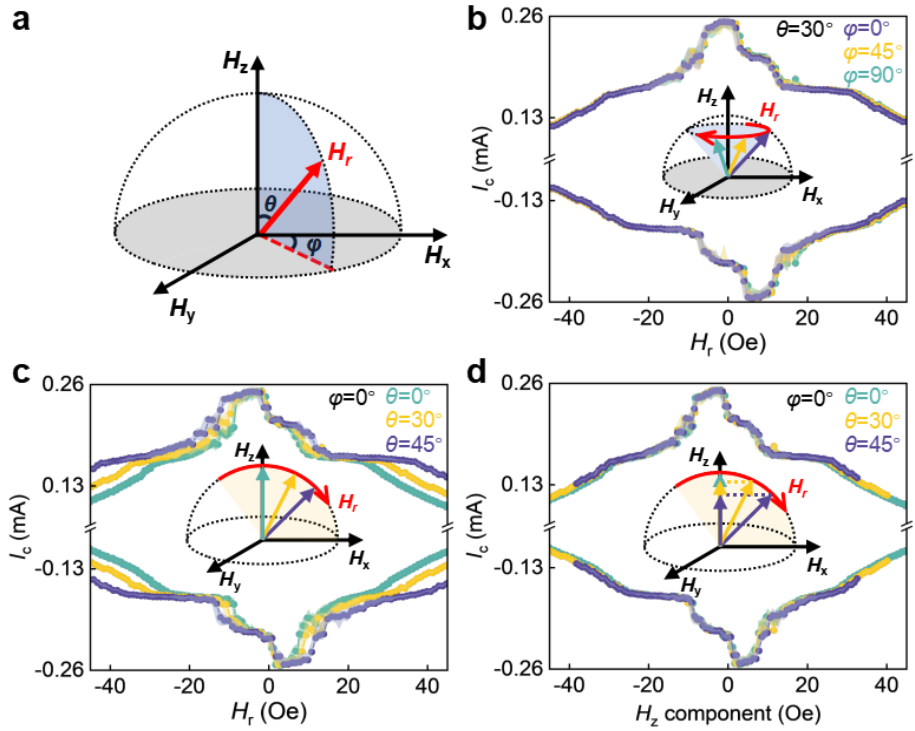

**Extended Data Fig. 1 | Angular dependence of critical currents at 80 K. a,** Definition of the spherical coordinate system used to describe magnetic field orientation, with polar ($\theta$) and azimuthal ($\varphi$) angles. **b,** Critical currents measured under three magnetic field directions: $\theta=30°$, $\varphi=0°$, 45°, and 90°, respectively. Colored curves correspond to different field directions indicated by arrows in the inset. Dots represent the averaged critical currents from 100 repeated measurements. **c,** Critical currents measured under magnetic fields confined to the *x–z* plane ($\varphi=0°$), for $\theta=0°$, 30°, and 45°. **d,** Critical currents plotted as a function of the out-of-plane field component ($H_z$), extracted from (**c**). The near-complete overlap of three curves demonstrates that the diode effect is primarily governed by the out-of-plane field component, with negligible contribution from the in-plane field.

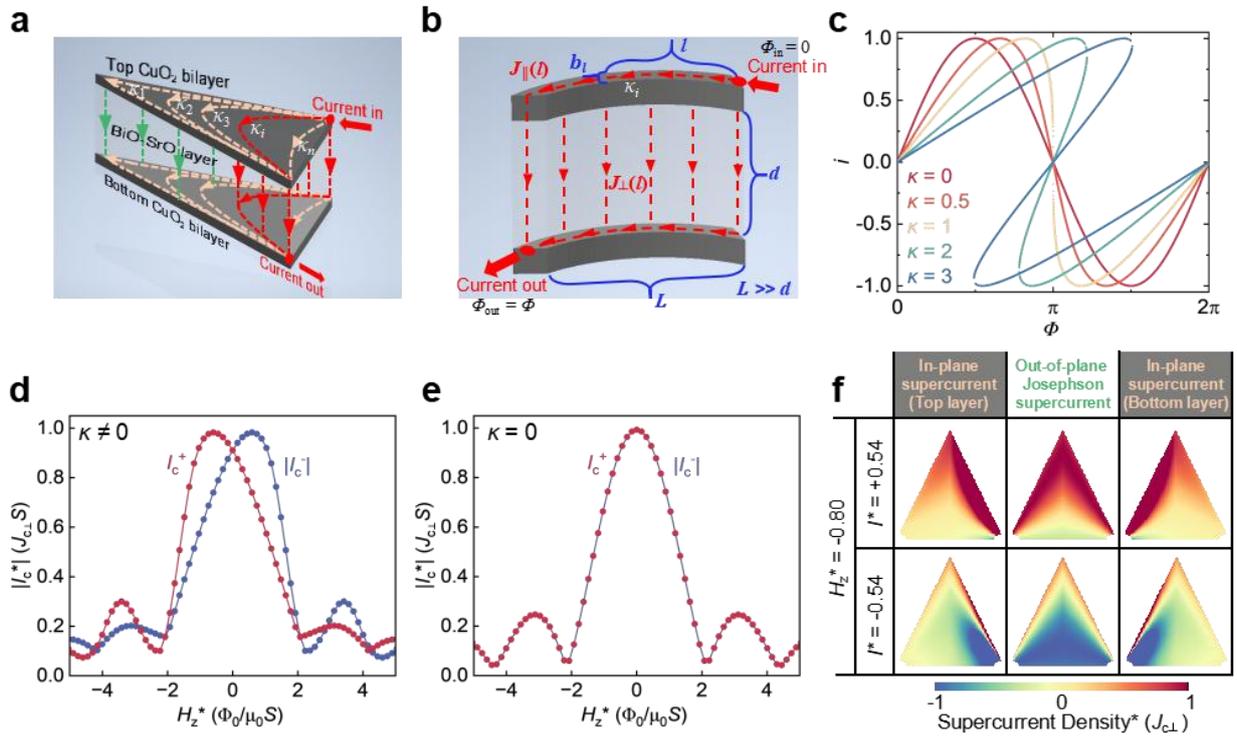

**Extended Data Fig. 2 | The current channel model and role of anharmonicity in nonreciprocity.**
**a,** Schematic of the theoretical framework, accounting for in-plane supercurrents in the superconducting $CuO_2$ bilayers (orange arrows) and the out-of-plane interlayer Josephson supercurrents (green arrows). A representative current channel is outlined with a dashed red contour, as shown in (**b**). The model comprises multiple current channels, each characterized by anharmonicity parameters $\kappa_1$, $\kappa_2$…$\kappa_n$, with the total current being the sum of the contributions from all individual channels. **b,** Schematic illustration of a current channel within an intrinsic Josephson junction. **c,** Anharmonic current-phase relation for the current channel, modulated by tuning the anharmonicity parameter $\kappa$. **d,** Theoretical field-dependent critical currents for $\kappa \neq 0$. Two curves exhibit clear nonreciprocity when a finite anharmonicity parameter $\kappa$ is introduced. **e,** Theoretical field-dependent critical currents for $\kappa=0$. The absence of Josephson diode behavior (perfect overlap of the critical currents for positive and negative bias) confirms that a nonzero $\kappa$ is essential for Josephson diodes. **f,** Spatial distributions of supercurrent densities with $\kappa=0$. Similar to the results in Fig. 2c, clear differences between positive and negative bias are seen, indicating time-reversal symmetry breaking under a finite magnetic field.

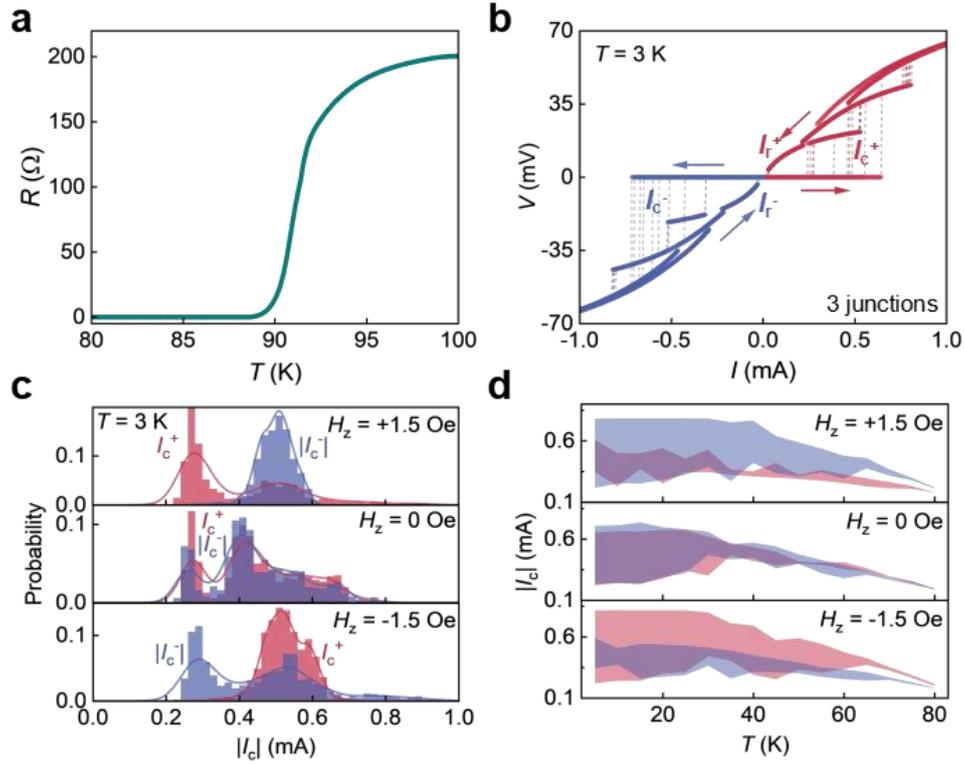

**Extended Data Fig. 3 | Transport properties of the intrinsic Josephson junction device. a,** Temperature dependence of resistance, which is measured under a 10 μA bias by subtracting voltages from opposite current polarities to eliminate offset. The superconducting transition temperature of the device is approximately 90 K. **b,** Current-voltage characteristics with ten repeated sweeps under cyclic bias current. Arrows denote sweep direction. Red and blue curves correspond to positive and negative current branches, respectively. The dashed line indicates the critical currents. The three voltage branches indicate the presence of three intrinsic Josephson junctions[42]. **c,** Statistical distributions of critical currents. Each panel shows the probability distribution of $I_c$ obtained from 1,000 repeated current-voltage measurements under different magnetic fields. **d,** Temperature dependence of critical current distributions. Shaded areas indicate the full range of critical current distributions (positive: red, negative: blue) as a function of temperature, obtained from 100 repeated current-voltage measurements.

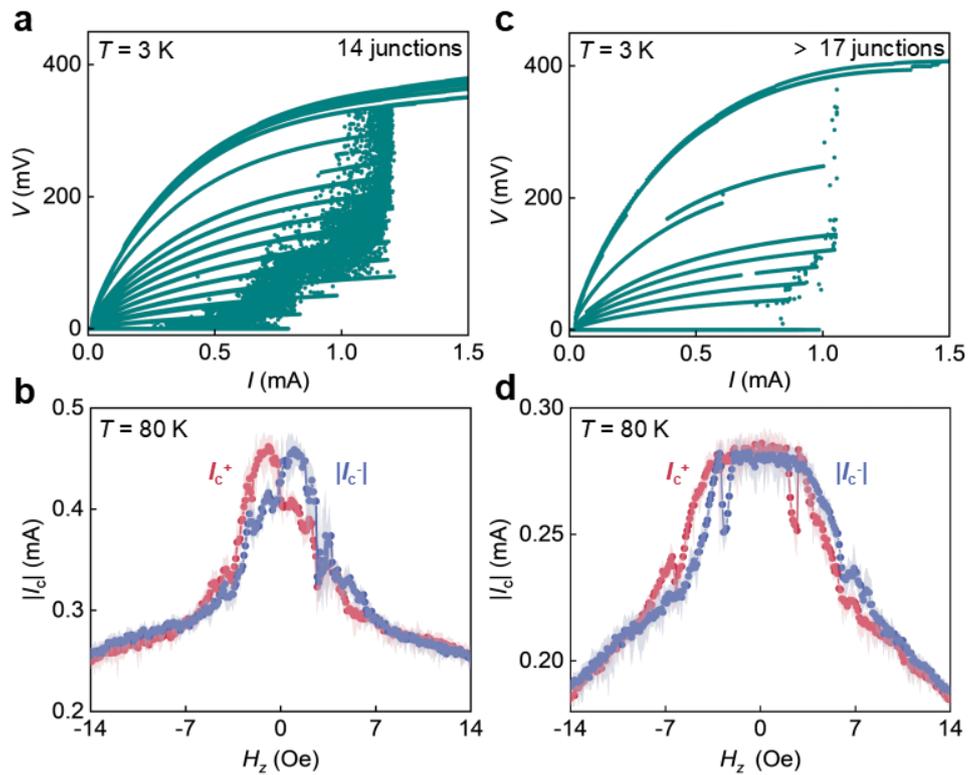

**Extended Data Fig. 4 | Multicycle transport measurements of two additional intrinsic Josephson junction devices. a and c** Current-voltage characteristics from 1,000 repeated measurements. The number of voltage branches reflects the number of junctions in each device. **b and d** Magnetic field dependence of critical currents corresponding to devices in (**a**) and (**c**), respectively. Solid dots indicate averaged values from 100 measurements, and shaded regions indicate distribution widths.

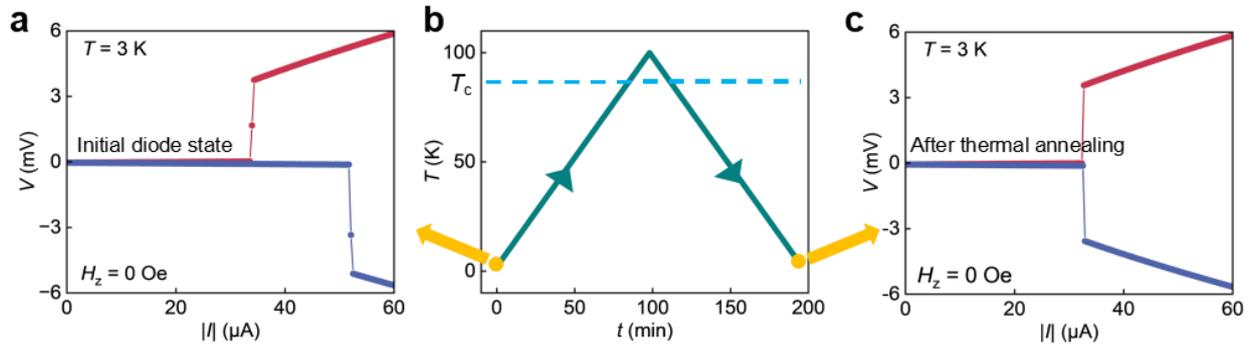

**Extended Data Fig. 5 | Thermal erasure of the zero-field Josephson diode effect. a,** Stable diode behavior observed at zero magnetic field. **b,** Schematic of thermal cycling: initial measurement in (**a**) at 3 K, heating up to 100 K, and re-cooling measurement in (**c**) at 3 K. The blue dashed line indicates the superconducting transition temperature ($T_c$) of BSCCO crystals. **c,** Post-annealing current-voltage characteristics. After the thermal cycling, the nonreciprocity vanishes at zero field, indicating that the zero-field Josephson diode state is erased by thermal treatment.

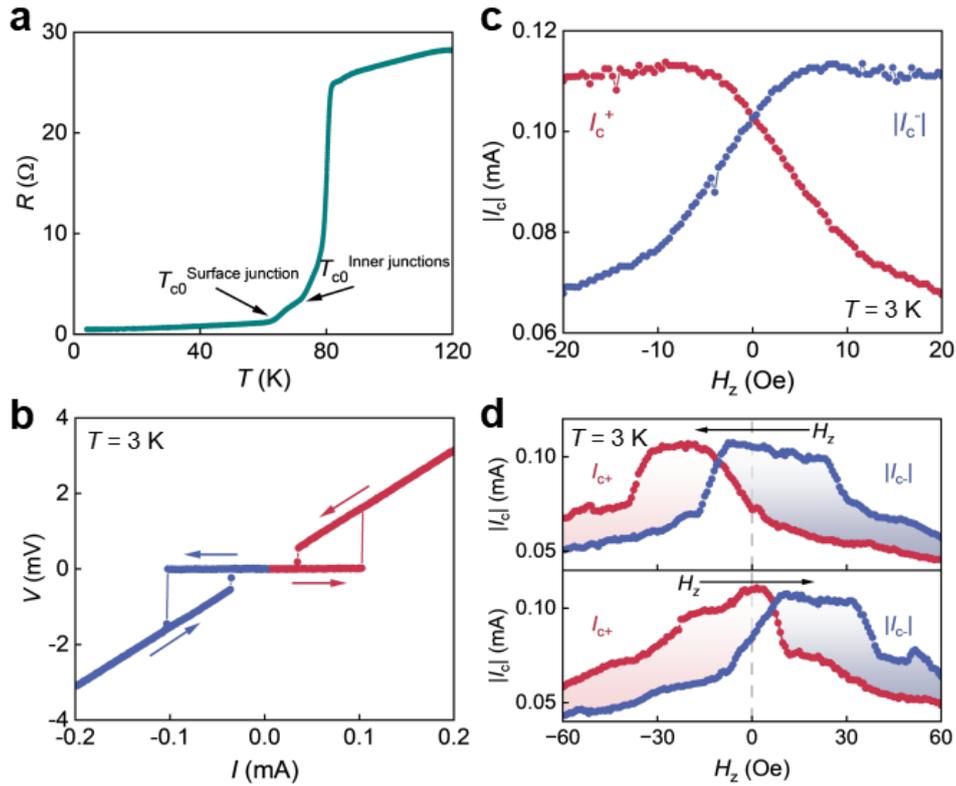

**Extended Data Fig. 6 | Transport properties of a selected surface intrinsic Josephson diode from the large array. a,** Temperature dependence of resistance. The bias current is 10 μA, and voltages were obtained by subtracting responses from positive and negative currents to eliminate voltage offsets. **b,** Current-voltage characteristics at zero field. Arrows indicate the sweep direction of bias currents. **c,** Magnetic field dependence of critical currents in the low-field region. The critical currents for reversed bias are symmetric at zero field. **d,** Magnetic field dependence of critical currents under high-field sweep conditions. The results reveal history-dependent critical currents ($I_c^+$ and $|I_c^-|$), with the zero-field difference highlighting the memory effect.

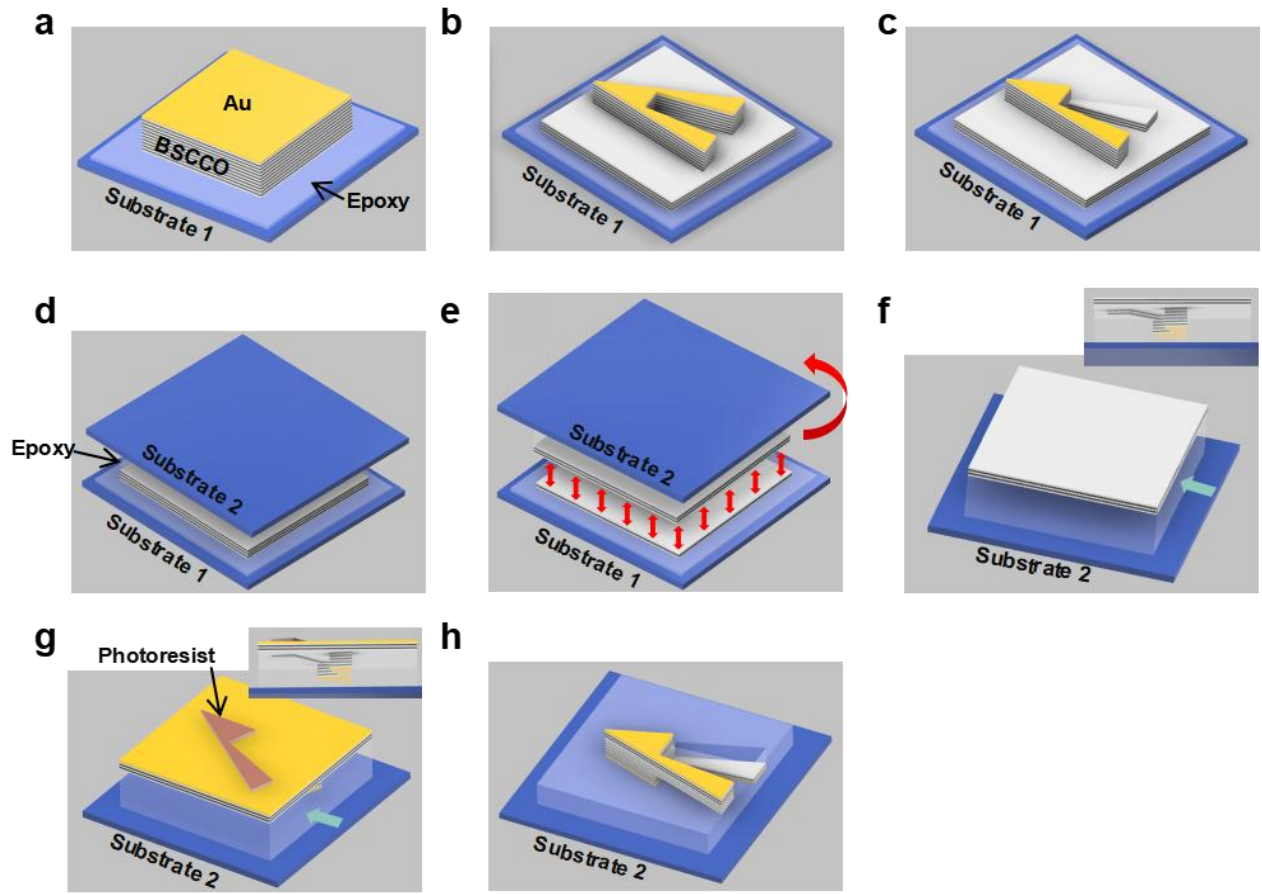

**Extended Data Fig. 7 | Fabrication procedure of intrinsic Josephson diodes. a,** Crystal mounting followed by gold film deposition. **b,** Patterning of the wedge-shaped crystal mesa. **c,** Defining the electrode on one side. **d,** Epoxy encapsulation and placement of the 2$^{nd}$ substrate. **e,** Splitting two substrates. **f,** Flip-transfer of the device onto the 2$^{nd}$ substrate. **g,** Deposition of gold film and photolithography to define the final electrode. **h,** Ion milling with precise control over the number of stacked intrinsic junctions.

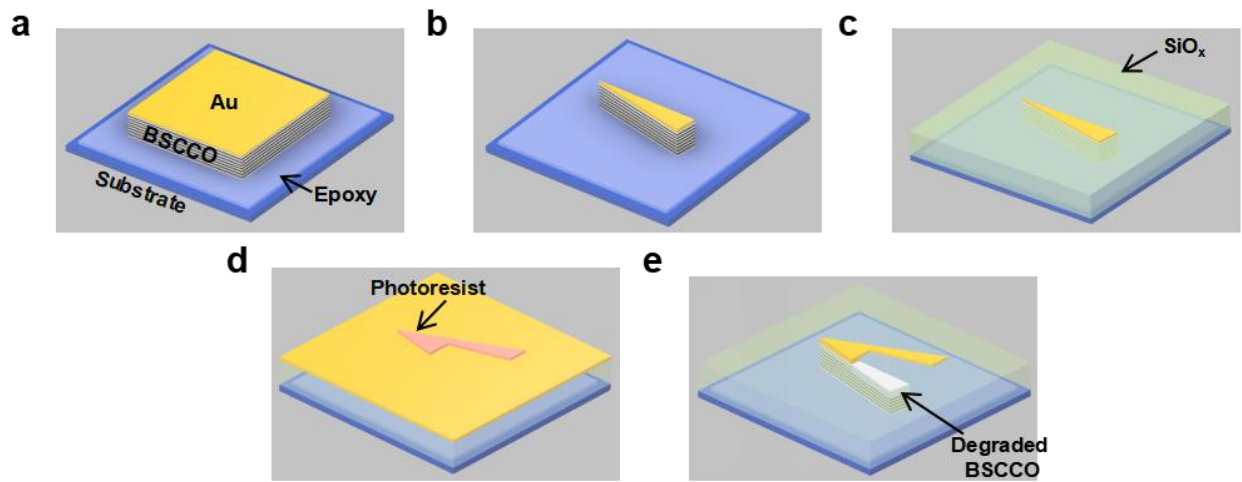

**Extended Data Fig. 8 | Fabrication procedure of the surface intrinsic Josephson diodes. a,** Crystal mounting followed by gold film deposition. **b,** Patterning of the wedge-shaped stack. **c,** Deposition of an insulating layer. **d,** Gold film deposition and photolithography. **e,** Wet etching of gold film to define electrodes.

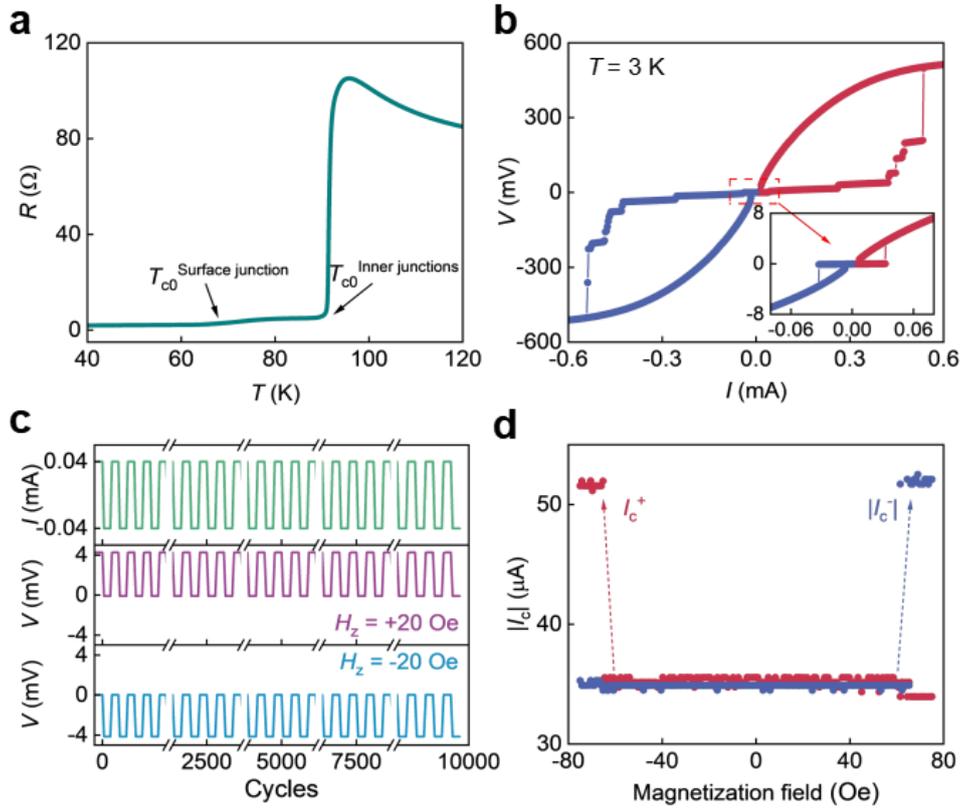

**Extended Data Fig. 9 | Transport properties of the surface intrinsic Josephson junction device.**
**a,** Temperature dependence of resistance. The bias current is 10 µA, and voltages were measured by subtracting the signals from positive and negative currents to eliminate offsets. Two distinct resistance kinks correspond to the superconducting transitions of stacked intrinsic Josephson junctions and the surface intrinsic Josephson junction, respectively. **b,** Full-range current-voltage characteristics. A voltage jump (~500 mV) indicates the presence of tens of intrinsic Josephson junctions in the device. The inset shows an enlarged view of the low-bias response, highlighting the characteristic from the surface intrinsic Josephson junction. **c,** Half-wave rectification at 3 K. Under square-wave current excitation, the device generates stable rectified voltage output under a magnetic of ± 20 Oe. Repeatable rectification is observed for over 10,000 cycles. **d,** Magnetization-programmed diode behavior. Critical currents were measured at zero field after applying different magnetization fields. Nonreciprocal zero-field critical currents ($I_c^+$: red, $|I_c^-|$: blue), emerge when the magnetization field exceeds ±60 Oe.

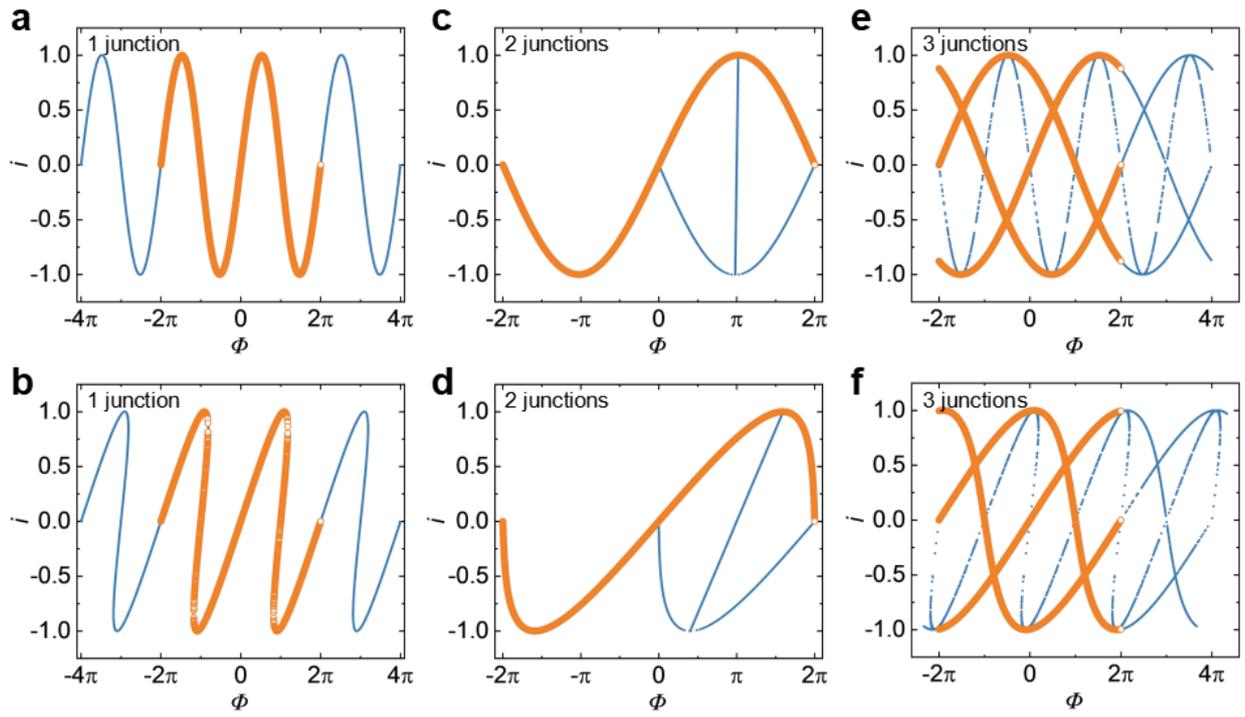

**Extended Data Fig. 10 | Comparison between numerical simulation and analytical model.** Numerically computed current-phase relations (blue) are compared with analytical results from Eq. (11) (orange) under zero magnetic field. The top and bottom panels correspond to small and large anharmonicity, respectively. Excellent agreement is observed between the two approaches.